\documentclass[a4paper,12pt]{article}
\usepackage[english]{babel}

\usepackage[dvips]{graphicx} 
\usepackage{amssymb} 
\usepackage{amsmath} 
\def\qqe#1{``#1''}

\begin{document}
\begin{sloppypar}

\pagestyle{empty}

\begin{center}
{\Large\textbf{A mathematical model of DNA degradation: Possible role of magnetic nanoparticles}}\\~ \\

V. N. Binhi
 
\end{center}

\noindent
General Physics Institute, Russian Academy of Sciences \\
38 Vavilova St., Moscow, 119991, Russian Federation. \\ 
E:  binhi@kapella.gpi.ru  

\vspace{10mm}

\noindent
\textbf{Abstract} 
\begin{quote}

A mathematical model of genome degradation is proposed that takes into account a variable rate of mutation and increasing number of cells in a developing human organism. The model explains known properties of cancer development, in particular, a synergism between different mutagens and an increased probability of cancer in the early years of life. An iteration equation is suggested that uses only a few model parameters and describes basic regularities observed in cancer onset. In the model context, relatively small chronic variations in the intracellular content of free radicals may markedly affect the probability of a cell to become a cancer cell. On the other hand, magnetic nanoparticles are shown to be an endogenous source of chronic magnetic exposure that increases the local concentration of free radicals. An enhanced level of leukaemia in early childhood is assumed to originate from magnetic nanoparticles located in hematopoietic stem cells.

\vspace{6mm}

\emph{Key words}: 
genome degradation, leukaemia, magnetosome, static magnetic field, hematopoietic stem cell, mathematical model, superparamagnetic nanoparticle, cancer risk factor, radical pair mechanism

\end{quote}

\section{Introduction}

\pagestyle{plain}

A number of studies have demonstrated that magnetic nanoparticles found in living tissues, including the human brain, are involved in biological reactions to magnetic field%
\footnote{Abbreviations: MF (magnetic field), HSC (hematopoietic stem cell), DNA (deoxyribonucleic acid), SD (single domain), SP (superparamagnetic), RP (radical pair).} %
exposures \cite{Johnsen.ea.2005}. However, the possible role of magnetic nanoparticles in molecular processes underlying cancer development has not yet been discussed.

It has been repeatedly demonstrated that organisms can biochemically precipitate minerals \cite{Lowenstam.1981} including magnetic minerals like magnetite Fe$_3$O$_4$ \cite{Lowenstam.1962,Blakemore.1975,frankel79}, maghemite $\gamma$-Fe$_2$O$_3$ \cite{Quintana.ea.2004}, and greigite Fe$_3$S$_4$ \cite{Matsunaga.ea.2000}. The presence of magnetic nanoparticles is well documented in many organisms \cite{Bazylinski.ea.2004}. They are also found in the human brain \cite{Kirschvink.ea.1992,Dobson.2002,Mikhaylova.ea.2004} and other human tissues \cite{Grassi.ea.1997}.

Membrane-enclosed crystals of magnetite are often called magnetosomes \cite{Balkwill.ea.1980}. Efforts were undertaken to identify genes required for magnetosome synthesis and arrangement. Some of the genes have been suggested to play a specific role \cite{Grunberg.ea.2001,Grunberg.ea.2004,Bazylinski.ea.2004,%
Matsunaga.ea.2005,Frankel.ea.2006,Fukuda.ea.2006}.

Magnetosomes are often assumed to underlie the observable biological effects from exposure to weak MFs \cite{Kirschvink.ea.2001}. The energy of a 100-nm magnetosome in the geomagnetic field is $\approx 24~ {k_{\rm B}{\rm T}} $. When exposed to an additional variable magnetic field $h$, the energy varies in about $(h/H_{\rm geo})24~{k_{\rm B}{\rm T}} $. If these changes exceed thermal fluctuations $\sim {k_{\rm B}{\rm T}} /2$, they can cause a biological response. This sets a natural constraint on the MF magnitude capable of affecting a biophysical or biochemical system appreciably: $h \gtrsim 1$--2~$\mu$T. Recent studies have shown that nonlinear stochastic dynamics of a magnetosome may be a basis for explaining biological effects from MFs of as low as 200~nT \cite{Binhi.ea.2005b}, and of biological magnetic navigation with a high accuracy \cite{Binhi.2006a}. 

The biological role of magnetic nanoparticles in higher organisms is not completely understood. The question is whether the consequences of their presence in biological tissues are limited only to magnetic orientation and possibly navigation by migrants, or can magnetic minerals take part in destructive processes in cells?

There are some indications that biological cells are not indifferent to magnetic nanoparticles. As was first noted in \cite{Binhi.ea.2005b}, MFs produced by magnetic nanoparticles are orders of magnitude greater than the geomagnetic field, and this can be an important endogenous source of chronic magnetic exposure facilitating free radical formation around the particles. Possible medical implications of human contamination by magnetic compounds are discussed in \cite{Purdey.2004}. Experimentally, ultrafine particles 12--14~nm in size were shown to be internalized by human monocyte cells and significantly increase, by 40--45\%, the release of free radicals \cite{Simko.ea.2006}. The severity of neurodegenerative diseases has been found to correlate with the amount of magnetite in the human brain \cite{Bartzokis.ea.1997,Hautot.ea.2003,Quintana.ea.2006}, which is explained by excessive toxic ferrous ions in brain. Formation of weakly charged ferric oxide nanoparticles in DNA complexes was discussed in \cite{Khomutov.2004.e} with regard to observation of broad ESR spectra in DNA preparations. 

It is assumed in the current article that some higher organisms have an abnormally increased content of magnetic nanoparticles in hematopoietic stem cells (HSCs). The content is low enough not to be detected by ordinary magnetic methods, since it is masked by the paramagnetic contribution of blood and the superparamagnetic contribution of ferritin. However, it is high enough to contribute to DNA lesions by chronically producing an additional amount of free radicals. HSCs are thought to be one of the primary places where acquired mutations can accumulate \cite{Huntly.ea.2005}.

The aim of this study is, first, to show within a mathematical model that even a small chronic change in the intracellular concentration of genotoxic agents such as free radicals can markedly enhance the rate of cancer incidence in the early stage of an organism's development and, second, to show that chronic changes in the free radical concentration can be induced by intracellular superparamagnetic nanoparticles.

Many mathematical models have been proposed to describe different aspects of cancer development: the evolutionary multistage model of cell proliferation in a changing environment \cite{Gatenby.ea.2003}, the multiscale model of key genes, cellular kinetics, and tissue dynamics \cite{Ribba.ea.2006}, the model involving the dynamics of gene inactivation with genetic instability \cite{Nowak.ea.2002,Nowak.ea.2004}, and others recently reviewed in \cite{Komarova.2005,Byrne.ea.2006}. Usually, the rate of mutation is considered to be a parameter. However, the rate of mutation may be a nontrivial function of the rate of DNA lesions, depending on cell cycle duration and the time needed to repair the lesions.

In this paper, a mathematical model is proposed that takes into account a variable rate of mutation. The model explains many known properties of cancer development: in particular, a synergism of chemical and radiation mutagens, and an increased probability of cancer in early years. It also provides an unobvious correlative link between the risk of getting some cancers in middle age and the time when acute exposure to an exogenous genotoxic factor took place. Numerical estimates are proposed for possible increases in the rate of cancer incidence from acute and relatively small chronic changes in intracellular free radical content.

Some recent experiments on the biological effects of static magnetic fields are summarized, average proper magnetic fields generated by superparamagnetic nanoparticles are calculated, and an idealized free radical reaction is analyzed to show that a magnetic nanoparticle can increase the rate of free radical formation in a cell by a few percent, which is enough to cause marked consequences in the rate of DNA degradation.

\section{DNA degradation model} 

DNA molecule is a sequence of its relatively short parts, or genes, each of which encodes a specialized protein. Since a DNA is a very long chain, it can be damaged in one or several places, and often this happens without loss of its ability to replicate. If damages, or mutations, are not repaired to the beginning of next cell division they may become permanent mutations inherited in daughter cells. Mutated genes may cause the production of abnormal proteins leading to diseases, in particular to cancers. 

A cancer develops in a process that involves mutations most likely in a few from about a hundred known cancer-related genes \cite{Sompayrac.2004}. 
These key genes, called oncogenes and tumor suppressor genes, take part in providing a dynamical balance between growth and death of cells of a given type. 

Various agents are known to induce mutations in living cells: many chemicals, aggressive free radicals, radioactive and ultraviolet radiations. Spontaneous mutations also happen. 

Normally, gene mutations are repaired by the DNA repair system. It includes a number of pathways through which an absolute majority of different damages in DNA may be recognized and repaired. The number of unrepaired mutations per year is only about a dozen in a cell. DNA repair system is a very perfect system that protects the body from diseases and untimely senescence. However, there are special genes encoding proteins, which regulate the DNA repair system, and these genes may also be corrupted by mutations. Then the rate of unrepaired mutations dramatically increases and a cancer becomes almost inevitable.

Most cancer researchers are focused on mutations in a small number of cancer-related genes as the primary damages which cause the subsequent malignant transformation of normal cells to cancer cells. There are other theories, in which some chromosomal abnormalities, like uneven number of chromosomes, are a root cause of every cancer \cite{Nowak.ea.2002,Gibbs.2004}. However, all agree that gene mutations are required. 

Evidently, mutations may accumulate only in proliferating cells. There are limited number of such cell types. HSCs play significant role in the human body. They are pluripotent cells: all other blood cells are made from adult stem cells in the bone marrow \cite{Levitt.ea.1995}. Like other blood cells, new HSCs are also originate from adult stem cells and gradually replace them in their functioning. It is the reason why HSCs are thought to be one of the primary places where acquired mutations can accumulate. The mutations are inherited in daughter stem cells, and this continuous self-renewal is accompanied by a consecutive DNA degradation. The gradual DNA degradation in HSCs is associated with the increased probability of leukaemias and, along with genome degradation in other cells, with natural ageing. If a hematopoietic stem cell became cancerous, it produces blood cells that further proliferate out of control and result for example in chronic or acute myeloleukaemias.

Below we will consider a mathematical model that illustrates main features of DNA degradation, to finally show why magnetic nanoparticles may be thought of as a cancer risk factor. In this model a cell experiences a set of $n$ consecutive renewals, which fix mutative damages accumulated during a given cell cycle. The probability $P_n$ of the cell to be a cancer cell at the end of cycle $n$ is a basic variable of the model that features the current state of the cell.

We will study first the process of the DNA degradation within the time interval $T$ of a single cell cycle. Because only about a hundred from 35000 genes in human genome are associated with cancers, and only a few of them, key genes, may relate to leukaemia, we consider one of these few genes and evaluate the probability $p$ that this gene is broken to the end of a cycle, if it was intact before the cycle. We will assume that a gene is a set of $k$ bits, and any damaged bit means the gene is damaged.

Let mutative DNA lesions be a Poisson process defined on a statistical ensemble of virtual identical bits. The probability for a lesion not to occur within the time interval $[0, t]$ is $\exp(-\alpha t)$, where $\alpha $ is the mean rate of lesions, both spontaneous and induced.

Let $t$ be the moment of a DNA lesion and $\tau $ be the time interval needed for the cell to repair any damage. In this idealized scheme, lesion is a random event and reparation is a deterministic process, which is necessary to occur if the cell has enough time before genome duplication, or mitosis, starts. Consequently, if $t <  T - \tau $, reparation occurs and no mutation appears in the gene. Oppositely, if $t> T - \tau $, the mutation will take place. 

The probability that a lesion will occur in the range $[t, t+{\rm d}t]$ is a product of two events: (i) the lesion had not occurred to the moment $t$ and (ii) it has occurred within the time interval ${\rm d}t$. It follows that the probability of the lesion occurring within $[t, t+{\rm d}t]$ equals $\exp(-\alpha t)\alpha {\rm d}t$. So the probability that a bit will be damaged in the range $[T-\tau , T]$, i.e. the probability of mutation, 
\[ p = \int_{T-\tau }^T \alpha e^{-\alpha t} {\rm d}t = e^{-\alpha T} \left( e^{\alpha \tau} -1 \right) . \]
Of course, the sum of the probabilities of the lesion to occur (i) before $T-\tau $, (ii) after $T$, and (iii) $p$ is equal to unity. Here it is assumed that more than one lesion in the same bit per cell cycle is an unlikely event.

\begin{figure}[t]
\[ \includegraphics[width=0.75\textwidth]{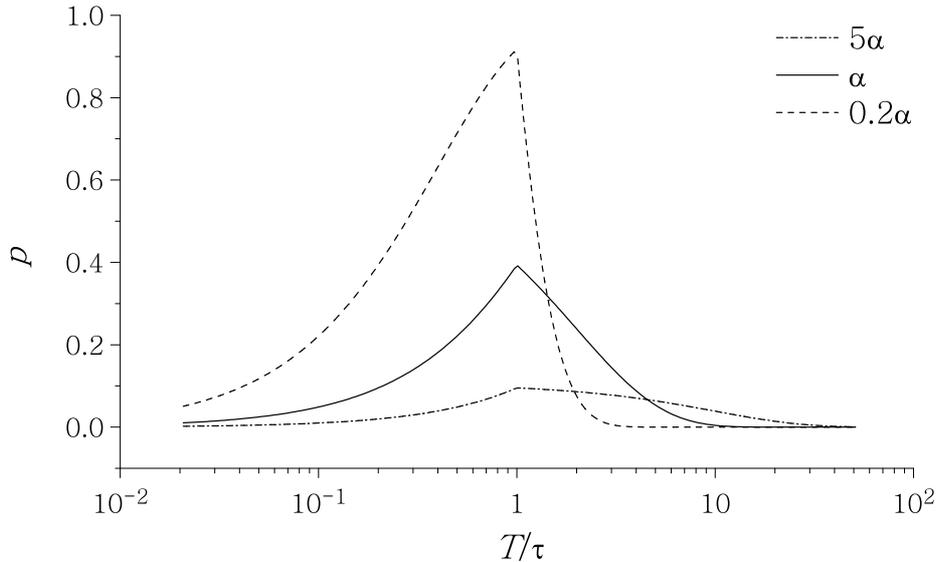} \]
\caption{\label{degrad-f01} {~~} {\protect \parbox[t]{100mm}{
The probability that a bit of a gene is damaged to the end of a cell cycle. To the left of $T/\tau=1$ the probability increases that the cell has enough time to repair a damage; to the right of $T/\tau =1 $ lesions have no time to occur. $\alpha $ is the mean rate of lesions.
} }}
\end{figure}

Next, we define that if the reparation time $\tau $ is greater than the cell cycle duration $T$, the probability $p$ no longer depends on $\tau $ and remains at that level where $\tau = T$, i.e. $1-\exp(-\alpha T)$. Then
\begin{equation}\label{degrad02} p(\alpha,\tau,T) = \left\{ \begin{array}{cl} e^{-\alpha T} \left( e^{\alpha \tau} -1 \right) ~, & \tau \leq T \\ 1-e^{-\alpha T} ~, & \tau > T ~.\\ \end{array} \right. \end{equation}
This function is plotted in the Fig.~\ref{degrad-f01}.

The probability that a gene of $k$ bits is undamaged to the end of a cell cycle is $(1- p)^k$, and that it is damaged $1- (1- p)^k$. 

Cells carrying the damaged genes cannot work properly. Most of such cells undergo apoptosis or the immune system will detect and kill them. However, a very small part of such mutant cells retain their functions and continue dividing. Let $u$ be the probability that a cell carrying a damaged gene will enter the next cell cycle. Then, an acquired mutation appear with the probability $\tilde{p}= u [1- (1- p)^k]$. The probability of no acquired mutation is $\tilde{q} = 1-\tilde{p}$ to the end of a cell cycle, and $Q_n = \tilde{q} ^N Q_{n-1}$ to the end of cycle $n$ in a population of $N$ cells. Actually, both $\tilde{q}$ and $N$ depend on $n$, and so $Q_n = \tilde{q}_n ^{N_n} Q_{n-1}$. Then the probability of the population to have an acquired mutation of a key gene in a cell to cycle $n$ is 
\begin{equation} \label{degrad03} P_n = 1-Q_n = 1- (1-P_{n-1}) (1-\tilde{p}_n)^{N_n} = 1- (1-P_{n-1}) \left\{ 1-u [1- (1- p_n)^k] \right\} ^{N_n}. \end{equation}
This quantity, which is associated with the probability of cancer, is a subject of the subsequent study.

As was said, the DNA repair system may be corrupted by mutations. In the model, there is the only parameter that describes the efficiency of the repair system, the reparation time $\tau $. The dependence of reparation efficiency on mutations means that $\tau $ is not a parameter but a variable $\tau_n$. We assume as a model approach that the repair system depends on mutations to different key genes in the same manner, irrespective of gene specificity. This enables to farther simplify the model: the only key gene will be studied, and the probability of it to be damaged reflects that of the cell genome as a whole. In this case, the variable $\tau_n$ depends on $P_n$. 

By definition, the value $P_n$ increases monotonously between 0 and 1. At $P_n =0$, the reparation time should be equal to its initial value $\tau (0) \equiv \nu $. At $P_n \rightarrow 1$, $\tau_n$ should tend to a maximum, for which $T$ is a suitable value, as was discussed above. In addition, $\tau_n$ should depend on the value of $P$ at the previous cell cycle. As a reasonable approximation, we take a linear dependence $\tau_n = \nu + P_{n-1} (T- \nu )$. Substituting this in (\ref{degrad02}) and using the designations as follows
\begin{equation}\label{degrad031}
a\equiv \alpha T~,~~~ b\equiv \alpha \nu ~,~~~ c\equiv a-b >0 , \end{equation} 
we write, for the case $\tau < T$, which takes place almost always except a special situation considered later,
\begin{equation} \label{deg0311} p_n = b+ c P_{n-1} +o(b) +o(c) , \end{equation} 
where $a\ll 1$, $b\ll 1$. In what follows, the mathematical indications of the accuracy of approximated equalities are omitted for convenience. Given this, we use the sign of exact equality keeping in mind that only terms linear in the small parameters $a$, $b$, $u$, and others are saved. Where these parameters appear as products like $akN$, which are not necessarily small, it will be specially indicated.

Substituting the latter equation in (\ref{degrad03}) gives an iteration equation
\begin{equation} \label{degrad033} P_n =  1- (1-P_{n-1}) \left\{ 1-u [1- (1- b - c P_{n-1})^k] \right\} ^{N_n}. \end{equation}

\begin{figure}[t]
\[ \includegraphics[width=0.75\textwidth]%
{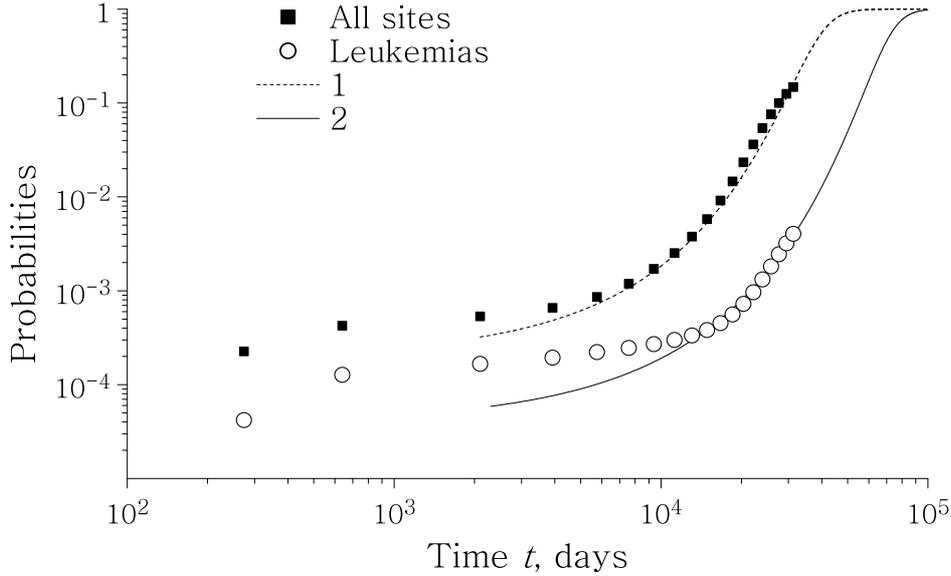} \]
\caption{\label{P-largetimes} {~~} {\protect \parbox[t]{100mm}{
Probabilities $P$ that a key gene mutation appears in a population of $N$ cells to the time $t$ calculated from equation (\ref{degrad033}) with $N=5\times 10^7$, curve 1, and $N=5\times 10^6$, curve 2. Points: cancer rates in the USA, all races, male and female, adapted from SEER 13 Registries for 1994--2003, http://seer.cancer.gov.
} }}
\end{figure}

In order to study the behavior of the probabilities at large times, we assume that the number of cells $N_n$ has reached its stationary value $N$. The results of computation of equation (\ref{degrad033}) are shown in Fig.~\ref{P-largetimes} along with epidemiological data on cancer rates in the USA. Calculations were made with $\alpha = 10^{-6}$ and $k=2\times 10^3$. For \qqe{leukaemias} case other parameters were $N=5\times 10^6$, $T=10$, $\nu =2\times 10^{-4}$, $u=1.6\times 10^{-8}$. For \qqe{all sites} case, $N=5\times 10^7$, $T=100$, $\nu =2\times 10^{-3}$, and $u=2.2 \times 10^{-9}$, all times in days and time variable $t=nT$. Parameter $\nu $ controls the slope of the curves in the inflection area. Parameter $u$, related to the immune and apoptotic processes, and $\nu $ were fitting parameters, and the product $\alpha u$ rather than $\alpha$ and $u$ separately was significant for calculations. Numerical values found from this fit are discussed below and used in what follows as reference values around which the quantities  of individual organisms may vary.

As is seen, there are three distinct areas of curves' behavior. It is slow quasistationary raise, an inflection area, and fast approaching to unity. On the whole, such behavior is associated with a slow-growing probability of cancer until the inevitable genomic degeneration against the background of senescence-related loss in the quality of genetic information.

\begin{figure}[t]
\[ \includegraphics[width=0.75\textwidth]%
{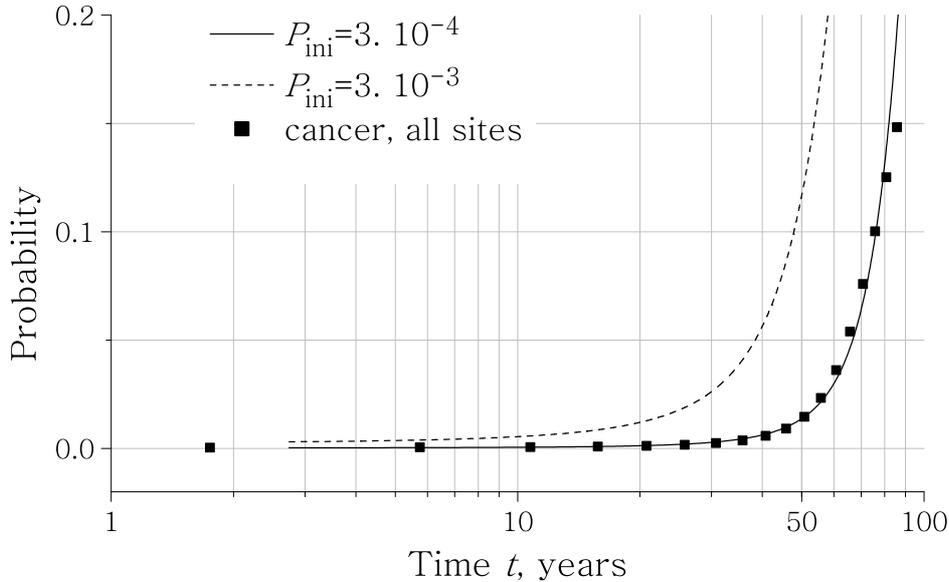} \]
\caption{\label{P-deflection} {~~} {\protect \parbox[t]{100mm}{
Shift of probability curve due to variation of the initial value of $P$ in iteration equation (\ref{degrad033}). Points: cancer rates in the USA, adapted from http://seer.cancer.gov.
} }}
\end{figure}

The time position of curves completely depends on the value of $P$ taken as initial value for the iteration process at large times, with all other parameters fixed. When $P_{\rm ini}$ is 10 times greater, probability of cancer starts to grow significantly by 30 years earlier, Fig.~\ref{P-deflection}, i.e. about 10-year shift per duplication of $P_{\rm ini}$. Formation of this value of $P$ at times about several years occurs in early stage of population development, along with acquired mutations. It is interesting to find an analytical solution of equation (\ref{degrad033}) in the range of relatively small values of $n$, since it is the interval where genetic mutations rapidly accumulate.

In this case $P_n\ll 1$ and equation (\ref{degrad033}) may be simplified. Based on digital values used for the parameters in the above calculations, one can see that $kb\ll 1$ and $(1-b -cP)^k = 1-kb -kcP$, then 
\[
P_n = 1- (1-P_{n-1}) \left\{ 1-u k[ b + c P_{n-1}] \right\} ^{N_n}. \] As well, $ukNc \ll 1$ that yields the next equation
\[ P_n = P_{n-1} + ukbN_n, \] which approximates (\ref{degrad033}) with an accuracy better than 1\%. Since the increment of $P_n$ at each iteration step is very small, it is convenient to reduce this iteration equation to a differential equation. Formally replacing $n$ with a continuous variable $\theta $ and $P_n$ with $P(\theta )$, we write 
${\rm d}P=ukbN^{^{}}{\rm d}\theta $. Since $b\equiv \alpha \nu  $ and all the parameters except $k$ may depend on $\theta $, the solution reads
\begin{equation} \label{degrad07} P(\theta ) = k\int_0^{\theta } \alpha \nu  u N ^{^{}} {\rm d}\theta, \end{equation} where for simplicity $\theta $ is used also as an integration variable. 
It is an important relation demonstrating the basic regularities observable in cancer development: 

(i) Genomic degeneration, or accumulation of acquired mutations, has a character of perfect integration. Consequently, the effects of exposures to even a short-term mutative factors will \emph{never abate} and will necessarily contribute to the overall degradation of the organism.

(ii) There are different channels by which the rate of genetic degradation might be influenced. It is $\alpha $ (rate of lesions), $\nu $ (rate of DNA repair), $u$ (immune responses and apoptosis), and $N$ (number of cells and cell cycle duration). 

(iii) External factors affecting genome through the channels of influence and doing it even relatively weakly but \emph{chronically} may significantly expedite the rapid growth of cancer risk.

(iv) Contributions of small changes in different channels are summed up arithmetically. However, there is a \emph{synergistic effect} of relatively strong simultaneous influences: the contributions are multiplied in this case. For example, short-term 10-fold simultaneous growth of any two parameters gives rise to 100 times greater degradation rate. Probably, free radicals are so dangerous as they affect simultaneously many channels.

The time variable $\theta $ in equation (\ref{degrad07}) is a dimensionless variable, a time measured in the units of cell cycle duration $T$. Generally, $T$ depends on time. It is well known that many cell types divide at a varying rate so that $T$ being minimal in the beginning then increases until it reaches a stationary value or even tends to infinity when cells stop dividing. This introduces an interesting feature concerned with the possibility that $T$ goes through the value of $\tau $ where the probability $p$ that a gene is damaged to the end of a cell cycle is maximal, Fig.~\ref{degrad-f01}.

In this case, the approximation $\tau < T$, used in (\ref{deg0311}), does not work and $p_n$, which now does not depend on $P$, in the initial phase where $P\ll 1$, should be written as
\[ p = \left\{ \begin{array}{cc} \alpha \nu & \nu \le T \\ \alpha T, & \nu > T \end{array} \right. , \] which follows from (\ref{degrad02}). Substituting it into (\ref{degrad03}) with $ukNb \ll 1$ we obtain instead of (\ref{degrad07}) the  equation
\[ P(\theta ) = k\int_0^{\theta } p(\alpha, \nu, T)  u N ^{^{}} {\rm d} \theta, \] where all the variables are functions of $\theta $. We will change variable $\theta $, i.e. dimensionless time expressed in $T$ units, to time $t$. As the cell cycle time $T$ changes with time $t$, the number of the cell cycles $n$ and as well $\theta $ completed to a moment $t$ will not be directly proportional to $t$:
\[ \theta(t)= \int_0^t \frac {{\rm d}t} {T(t)} . \] Consequently, since the function under the following integral does not depend on $t$ explicitly, 
\[ P= k\int_0^{\theta } pu N ^{^{}} {\rm d}\theta = k\int_0^t p  u N \frac {{\rm d} \theta } {{\rm d}t} ^{^{}} {\rm d}t  , \] or
\begin{equation} \label{deg081} 
P(t) = k\int_0^t \frac { p(\alpha, \nu, T)  u N } {T}  ^{^{}} {\rm d}t , \end{equation} where all the variables are now functions of physical time $t$. To analyze this equation, some ideas on the character of functions $N(t)$, $T(t)$, and $u(t)$ should be formulated.

\begin{figure}[t]
\[ \includegraphics[width=0.75\textwidth]%
{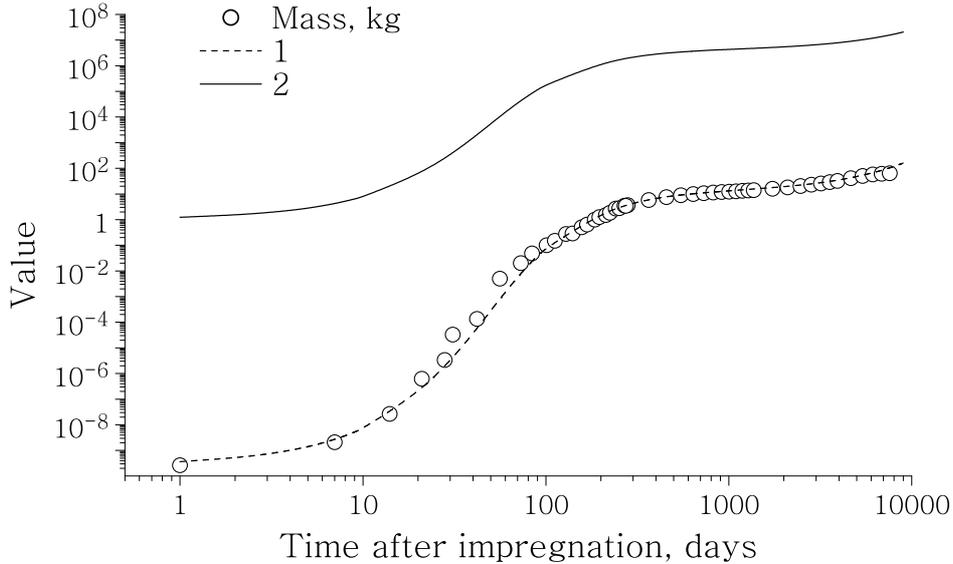} \]
\caption{\label{deg-NumCells} {~~} {\protect \parbox[t]{100mm}{
Human organism mass and the number of HSCs. Points: treatment of averaged data on masses of the human embryo, fetus, and body \protect\cite{Human.2006}. Lines: (1) approximation of the time dependence of the mass $M(t)$, (2) scaling of the approximation curve $M(t)$.
} }}
\end{figure}

There are no age-specific data on the quantity of HSCs in human body. One might assume it to be, in a sense, proportionate to the average mass of a growing organism. Fig.~\ref{deg-NumCells} demonstrates pooled data on the masses of human embryo, fetus, and body over time after impregnation. The data were approximated by a smooth curve 1, which is denoted $M(t)$, then scaled linearly in log scale\footnote{
Log-linear scaling is the simplest nonlinear scaling that transpose variable $y$ to $y'$, values $y_1$ to $y'_1$ and $y_2$ to $y'_2$ so that this transposition is a linear one of a logarithm of $y$: $y'=  y'_1 \exp \left[ \ln 
\left( y/y_1 \right) \ln \left( y'_2/y'_1 \right) \ln^{-1} \left( y_2 /y_1 \right) \right]$.   
} so as to move the first point to unit and last points to about $5\times 10^6$, known estimate of the number of HSCs in adults, curve 2. This dependence $N(t)$ is used in further calculations. 

\begin{figure}[t]
\[ \includegraphics[width=0.75\textwidth]%
{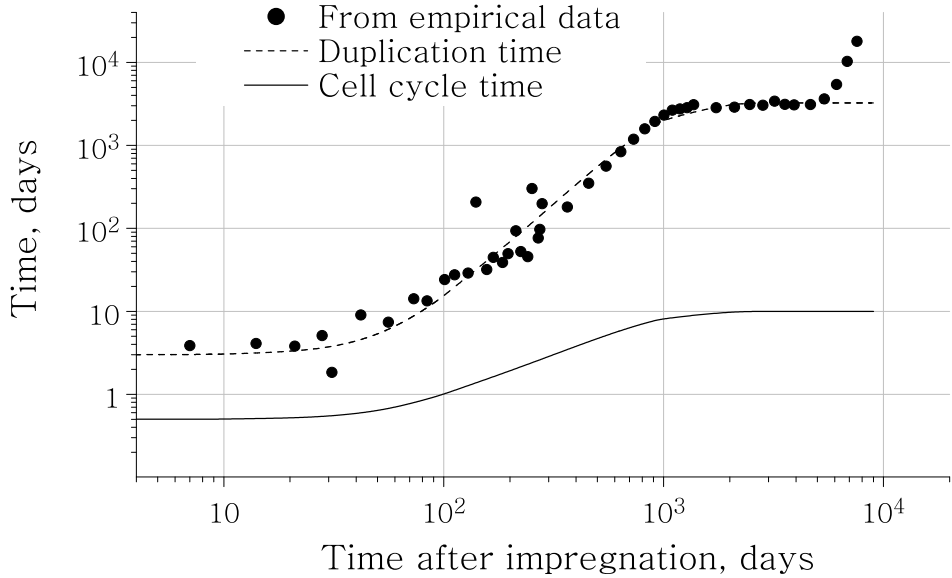} \]
\caption{\label{deg-CellCycle} {~~} {\protect \parbox[t]{100mm}{
Time-dependence of the cell cycle time $T$. Treatment of averaged data on masses of the human embryo, fetus, and body \protect\cite{Human.2006}. 
} }}
\end{figure}

Time-dependence of the cell cycle time $T(t)$ in humans may also be estimated from the age-dependence of the averaged human mass $M(t)$. The duplication time $T_{\rm d}$ at any moment of time follows from an obvious relation $M(t) + M'_t(t) T_{\rm d}(t) = 2 M(t)$. Then,
\begin{equation} \label{deg091} T_{\rm d}(t) = \frac {M(t)} {M'_t (t)} ~. \end{equation}
As shown in Fig.~\ref{deg-CellCycle}, this function, computed from previously gained curve $M(t)$, well approximates points calculated from the empirical data on mass of the human organism. This calculation has been made based on a discrete analog of (\ref{deg091}).

Mass duplication time $T_{\rm d}$ is not the cell cycle time $T$, though it directly relates to the latter. If cells of the human organism were dividing without death, $T$ would  be equal to $T_{\rm d}$, of course within a rough idealization in which a human organism is an ensemble of identical cells. However, one could assume that the regularity, which governs the exponential increase of the cell cycle duration, is approximately the same as for the mass duplication time. Then, a log-linear conversion would be a reasonable procedure to obtain $T(t)$ from $T_{\rm d}(t)$. The result of such conversion with 0.5~d and 10~d as initial and final values for $T(t)$ is shown in Fig.~\ref{deg-CellCycle}.

As an illustrative example only, we will take this function $T(t)$ to study the degradation of HSCs, which are assumed to follow this dependence in early stage of the development of an organism. So, both functions $N(t)$, number of cells in growing population, and $T(t)$, cell cycle duration, are built on the tabulated dependence $M(t)$ and fixed in the following calculations. This decreases the number of parameters, which is necessarily great in a model of such complex phenomena as cancer. On the one hand, even in this case we use initial and final values for $N$ and $T$. On the other hand, these values are natural for leukemic cells, and the semiquantitative results of the model are not sensitive to reasonable variations of these values.

\begin{figure}[t]
\[  \includegraphics[width=0.75\textwidth]%
{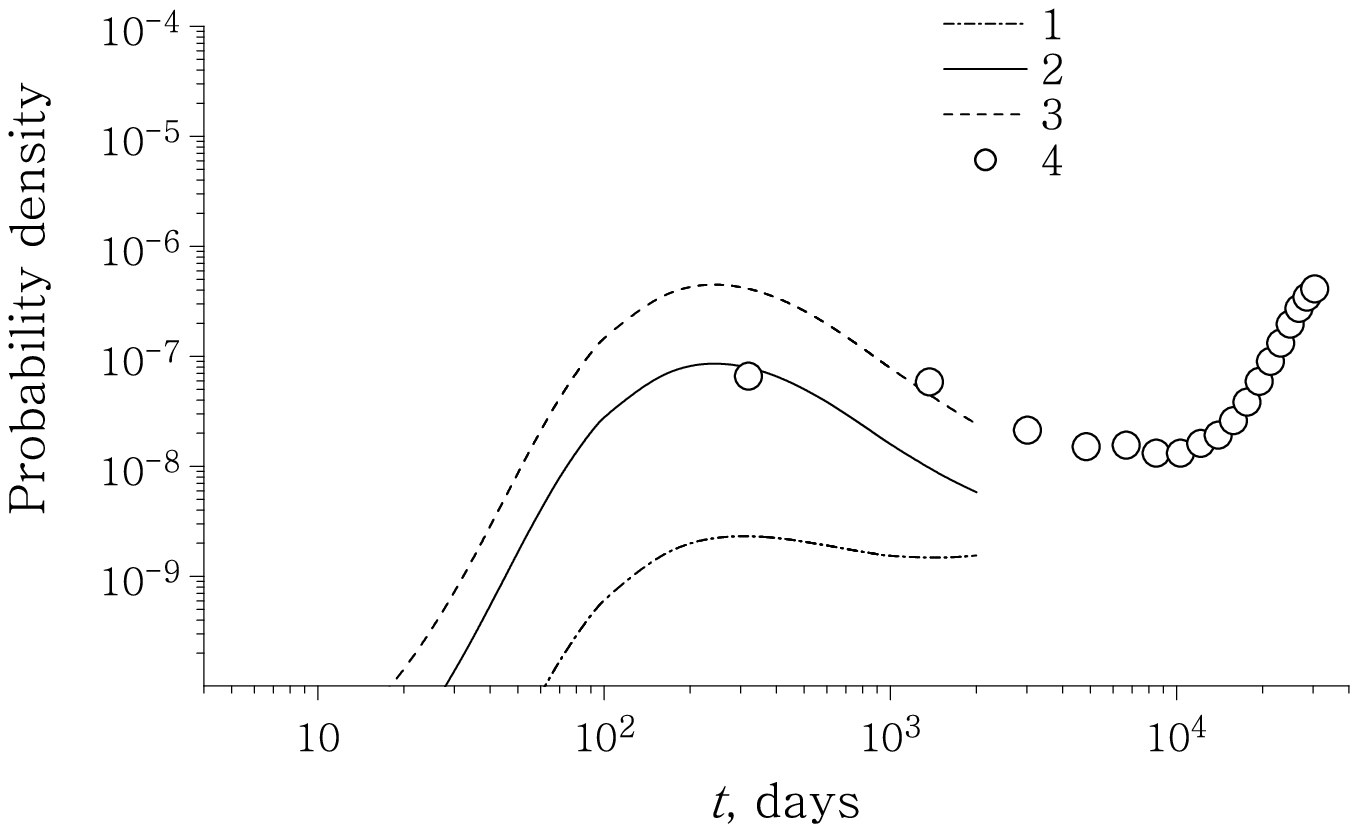} \]
\caption{\label{deg-Max} {~~} {\protect \parbox[t]{100mm}{
Leukemia probability density ${\rm d}P/{\rm d}t$ calculated from (\ref{deg081}): 1, $\nu =2\times 10^{-4}$ and $u =1.6\times 10^{-8}$ as in Fig.~\ref{P-largetimes}; 2, $u_0=7.6\times 10^{-7}$; 3, $u_0=4\times 10^{-6}$, explanation in the text. 
Age-specific leukaemia incidence rates in the USA, 4, adapted from  http://seer.cancer.gov.
} }}
\end{figure}

Two functions, $\nu (t)$ and $u(t)$, in (\ref{deg081}) remain to be modelled to calculate probability of leukaemias. First, we note that even taking these functions constant gives rise to a peak in ${\rm d}P/{\rm d}t$, due to an interplay between the functions $N(t)$ and $T(t)$, curve 1 on Fig.~\ref{deg-Max}. This might be associated with the observable growth of leukaemia incidence in early period of an organism's development. Position of the peak, some left of the empirical peak, depends on the tabulated curve $N(t)$, precisely on the position on its rapid growth, and less sensitive to other parameters. This mismatch is not principal. Maximum peak value will be studied then. 

At constant $\nu $ and $u$, the peak value is more than 10 times less than that of the empirical peak. This makes one think of another factors affecting the value of the local-in-time peak. The fact is that the peak value significantly depends on the character of $u(t)$ and $\nu(t)$ time-dependences. 

According to medical knowledge about the immune system and DNA repair system their efficiencies grow in time in the early phase of an organism's development. On the whole, the immune system monotonously unfolds as cell differentiation occurs. The immune system gets stronger as a lot of previously unknown alien dangerous proteins become known. Apparently, in a similar way, the DNA repair system develops gradually as a lot of necessary protective proteins are expressed from specific genes. So the assumption, which is based on basic medical knowledge, is that the functions $u(t)$ and $\nu(t)$ monotonously fall from an initial values to their final values, which means the efficiency of the immune and repair system grow correspondingly. 

The final values of $u(t)$ and $\nu(t)$ have been already determined from the comparison with the empirical data, Fig.~\ref{P-largetimes}, within the frame of the model approach. Assuming that DNA repair system develops much faster than the immune system, we will take $\nu $ as a constant and focus on the contribution to ${\rm d}P/{\rm d}t$ from the immune system. If it were not so, synergistic effects might be great. It is convenient to model the function $u(t)$ as a logistic curve $u(t)= u_1 + (u_0 - u_1)/[1 + (t/t_0)^s] $ with $u_1=1.6\times 10^{-5}$, $u_0 =7.6\times 10^{-4}$, $s=2$, and $t_0 =500$~d, which gives a good fit, Fig.~\ref{deg-Max} curve 2. The changes in the peak value are shown with varying $u_0$, curves 1 and 3. Exponent $s$ regulates the peak width. The peak value also changes with parameter $t_0$, but in a lesser extent. If to designate the maximum peak value $\rho \equiv {\rm max}({\rm d}P/{\rm d}t)$ and define a relative sensitivity to varying a parameter $x$ by the formula $S\equiv ({x}/{\rho }) ({\partial \rho} /{\partial x})$, then the sensitivity to parameters $u_0$ and $\alpha $ is close to unity, and that to $t_0$ is about 0.37.

Some discrepancy between the model and empirical data in the position of the peak could not be eliminated by  varying model parameters in acceptable intervals. Most likely, the delay of empirical points vs calculated curve originates from the fact that it takes a few months in order that a single cancer cell, for which probability $P$ is derived, would generate a detectable population of cancer cells. 

It would be interesting to examine how variations in the parameters, caused by chronic detrimental factors, like an enhanced free radical concentration, could affect the value of the peak in the probability density of cancer ${\rm d}P/{\rm d}t$. However, it is hardly possible. 

Excessive free radicals affect not only the current values of the parameters, but the course of their formation. It means it would be impossible to change the value of a parameter at a moment $t$ without significant changing all the subsequent time dependence of that parameter.
Only the rate of lesions could change directly with concentration of chronically acting genotoxic factors. The other parameters, $\nu $, $u$, $N$, and $T$ are systemic parameters. They describe the state of corresponding systems: the rate of DNA repair system, perfection of the immune system, the number of cells in population, and the averaged cell cycle duration. Each of these nonlinear functions depends on the prehistory of other systems as well. For example, the development of the immune system,  the growth of the number of cells, and changes in the averaged cell cycle duration are evidently coupled into interacting processes. This makes it difficult to estimate quantitatively the consequences of a chronic extra content of free radicals, except a special case below. 

Clear qualitative results for the case of chronic factors may be obtained from analyzing equation (\ref{deg081}). From (\ref{deg081}) follows that the time derivative of $P$, associated with age-specific cancer incidence rates, depend on several functions:
\[ 
\frac {{\rm d}P} {{\rm d}t}= kp(\alpha, \nu, T)uN/T. \] It can be shown that the peak in ${\rm d}P /{\rm d}t$
on Fig.~\ref{deg-Max} is formed mainly by two basic, opposite tendencies. Initial increase in ${\rm d}P /{\rm d}t$ is due to rapid exponential growth of the number of cells $N$. This is opposed by decreasing $u$ (growing perfection of the immune system) which finally prevails over the former tendency so that the peak is forming. The peak value strongly depends on whether the development of the immune system keeps pace with the growing number of cells. If it has not enough time or motivation to develop, a significant increase in cancer probability may happen. Two strategies are evident to minimize the risk: one strategy is to promote formation of the immune system as soon as possible, the other one is to slow down the rapid growth of the mass of an organism.

Unlike the effects of chronic factors, effects of short-term exposures to toxic factors may be easily calculated within the frames of this model. Consider again equation (\ref{deg081}), from which follows 
\begin{equation} \label{deg0101} 
P(t) = \int_0^{t} \frac {{\rm d}P} {{\rm d}t} ^{^{}} {\rm d}t . \end{equation} 
Consider the value of $P$ at a moment $t'$ preceding the phase of the rapid growth of $P$, Fig.~\ref{P-deflection}. As was shown, $P(t')$ determines at which time the subsequent growth of $P$ to dangerous values takes place. Time derivative ${\rm d}P/{\rm d}t$ is directly proportional to any of the parameters capable of a sudden short-term change, $\alpha $, $\nu $, and $u$. Let $x$ be one of them, and $\delta x$ be a short-term variation of it, localized at a moment $t$. Then the probability of a cancer to a moment $t'$ and the increment $\delta P$ are equal to 
\begin{equation} \label{deg092} P(t') = \int_0^{t'} \frac {x +\delta x} {x} \frac {{\rm d}P} {{\rm d}t} ^{^{}} {\rm d}t ,~~~ \delta P  = \frac {{\rm d}P (t) } {{\rm d}t} \int_t^{t+\delta t} \frac {\delta x} {x} ^{^{}} {\rm d}t, \end{equation}
because time derivative of $P$ practically remains unchanged during the time interval $\delta t$ within which $\delta x $ differs from zero. Evidently, $\delta P$ is the same for any $t' >t +\delta t$. Quantitatively, increment $\delta P$ may be estimated if the data are known about the relative growth of a risk factor and about the age-specific rates of the cancer incidence. 

An important qualitative result: The long-term effect of a short-term exposure varies directly with the value of age-specific rate of cancer incidence on the moment of exposure. Leukaemia and many other cancers show enhanced incidence in early childhood. For example, based on the empirical data displayed on Fig.~\ref{deg-Max}, one could conclude the following: medical x-ray procedures in the first few years are an order more hazardous, in terms of cancer probability in old years, than those carried out in a later period. To our knowledge, it is the first mathematical observation of the correlation between elevated risk of these forms of cancer in old years and exposure to genotoxic factors precisely in the first years of life.

Returning to chronic exposures, one could suggest a very rough estimate for the case when chronic variations $\delta x(t)$ of the parameters $x(t)$ are small and change linearly with $x$, i.e. $\delta x \propto x(t)$. In this case, the value $\delta x(t)/x$ does not depend on time. Then, conversions like in (\ref{deg092}) give rise to another relation for the increment $\delta P$, which now depends on the time $t'$ at which it is taken:
\[ \delta P(t') = \int_0^{t'} \frac {\delta x} {x} \frac {{\rm d}P} {{\rm d}t} ^{^{}} {\rm d}t = \frac {\delta x} {x} P(t') ,\] i.e. $\delta P {}^{}/P = \delta x {}^{}/x$. Indeed, there are serious epidemiological and empirical data, as well as rationale, for a linear dose-response curve for radiation-induced cancers, particularly
at low doses \cite{Preston.2003}.

In the absence of any quantitative data on how small free radical variations could affect the different channels, that is, the rate of lesions and repairs, perfection of the immune system, and cell cycle durations, one might only assume, on the whole, in average, and as a very rough approximation, a linear dependence between corresponding relative values. That is, for example, a chronic 1\%-change in the concentration of aggressive molecules causes 1\%-change in each of the above channels. Contributions of small changes in different channels are summed up in $P$ arithmetically, as was shown above. Moreover, since, on the whole, excessive free radicals do increase the rate of lesions, retard the DNA repair, impede the immune system to operate duly, and probably as a long-term systemic response to a danger stimulus, increase the size of population $N$ and decrease cell cycle $T$, then, according to (\ref{degrad07}) and (\ref{deg081}), changes in all channels should be summed with the same plus sign. Consequently, there are grounds, though qualitative, to think that a small but chronically acting relative elevation of free radical concentration results in at least a few-fold greater relative growth of cancer probability $P$.

Evidently, a 1\%-increase in the probability of cancer in age 5--10 years will inevitably shift the curve $P(t)$ up by the same 1{\%}, even if no increase happens in the rate of genotoxic factors later and all the essential parameters remain the same, Fig.~\ref{P-deflection}. It might seems to be an insignificant change. However, it means also that the curve $P(t)$ moves left by about one month, which should be considered as a direct influence of $P$-level in early childhood on the lifetime. Therefore, analysis of possible physical reasons, which can increase probability of cancer in childhood, is well grounded. One of the possible reasons, not addressed in the literature until now, is magnetic nanoparticles chronically exposing nearby molecules to rather strong static MFs, which in turn promote formation of free radicals.

\section{Magnetic nanoparticles}

The agreement between the above model calculations and observable data is rather conditional, since many assumptions have been made on the character of key dependences. However, the model provides a possibility to study with greater certainty how different external factors might affect the growth of cancer probability. 

Among factors capable of significantly promoting cancers are well known ionizing and ultraviolet radiations, some mutagenic chemicals related to tobacco use, environmental pollution, food contaminants, and some infectious agents. Other known risk factors are social factors: alcohol, obesity, unhealthy diet and physical inactivity, or factors intrinsic to an individual: age, sex, genotype and heredity.

Last decades have shown that this list of risk factors might be widened. Recent epidemiological studies revealed a possible cancer risk of weak extremely-low-frequency electromagnetic fields generated mainly by power lines and various home electrical appliances. There is a correlation between the intensity of background magnetic fields and the rate of childhood leukaemia incidence. But the correlation is observed not always and another part of epidemiological studies show no correlation. In spite of available controversy, the International Agency for Research on Cancer has classified magnetic fields as possible human
carcinogens. A possible link between weak magnetic fields and cancer is hotly debated in several recent meta-analyses \cite{Ahlbom.ea.2000,Forssen.ea.2000,%
Wartenberg.2001,Greenland.ea.2005} and reviews \cite{Kheifets.ea.2005,Morgan.ea.2005,Elwood.2006}. At the same time, the nature of the processes underlying this link remains unclear. In general, biological effects of electromagnetic fields, which do not appreciably heat biological tissues, look somewhat contradictory from the physical viewpoint, and there are no recognized physical mechanisms for such effects yet. 

A lot of hypothetical mechanisms have been suggested to explain biological effects of weak extremely-low-frequency MFs. A brief review of the mechanisms may be found in \cite{Binhi.ea.2003a.e} and the detailed examination in \cite{Binhi.2002}. Most researchers often discuss the following hypothetical physical targets for MF action in magnetobiological phenomena: (i) iron-bearing magnetic nanoparticles growing in biological tissues, (ii) spin-correlated radical pairs, in some biochemical reactions, interacting with magnetic field by their spin magnetic moments, (iii) long-lived rotational states of some molecules inside protein structures, which interact with MF by their orbital magnetic moments.

The basic problem is that the interaction energy of biologically active molecules and the MF at the geomagnetic level is very small. It is by many orders of magnitude smaller than the energy of thermal fluctuations ${k_{\rm B}{\rm T}}  \approx 4\times 10^{-14}$\,erg at physiological temperatures: $\mu H \ll {k_{\rm B}{\rm T}}  $, where $\mu $ is the molecular magnetic moment and T is the absolute temperature, which is written in Roman to distinguish it from the cell cycle time $T$. Obviously, magnetic effects cannot exist here! It is the most heated argument raised by the opponents of the idea that weak MFs, on the order of the geomagnetic field and lesser, can affect organisms. At the same time, an ever-growing number of experimental evidences demonstrating biological effects of weak MFs require further attentive and careful theoretical study.

Apparently, to explain observed biological effects of weak extremely-low-frequency magnetic fields one needs to equalize the inequality $\mu H \ll {k_{\rm B}{\rm T}}  $: either the magnetic moment $\mu $ of a suggested target in an organism should be sufficiently large, or the effective temperature $T$ of the target should be sufficiently small. The former possibility is used in mechanisms based on magnetic nanoparticles \cite{Kirschvink.ea.1981} found in tissues of many organisms including human brain tissues \cite{Kirschvink.ea.1992b,Mikhaylova.ea.2004}. Magnetic energy of such particles may exceed ${k_{\rm B}{\rm T}}  $ by many times and cause a biological response. In a recent work \cite{Binhi.ea.2005b}, as small MFs as of $200$~nT were shown to significantly change the nanoparticles' dynamics.

On the other hand, the above inequality relies on the implicit assumption that a target is in thermal equilibrium with the surrounding medium. Evidently, one could overcome the \qqe{kT problem} also having found possible targets whose effective temperature differs from that of the medium, so that the inequality would no longer be valid. Hence, no fundamental limitations would be placed on the possibility of observing biological effects of weak MFs that interact with such targets.  

Reactions involving free radical pairs give a clear example of the case where the inequality fails. Magnetic processes based on spin dynamics of the radicals develop so quickly that the thermodynamic equilibrium has no time to be established. This means spins move coherently and no temperature of spins exists within these small time intervals, for which the term \qqe{spin lifetime} is used. Another example where the inequality fails is the molecular gyroscope model \cite{Binhi.ea.2002b}. In this model, a small biologically active molecule bound within a protein cavity is well isolated from the surrounding thermal perturbations. Interacting with the MF, it can coherently rotate for a long time. This makes the notion of temperature inapplicable to this type of molecular targets as well. 

As is seen, there are different ways to overcome the kT problem, at least at the conceptual level. Among them, the most certain are mechanisms involving magnetic nanoparticles as a primary target for MFs \cite{Diebel.ea.2000}. A lot of studies demonstrate their presence in living tissues, on the one hand, and their involvement into biological reactions, on the other hand, \cite{Johnsen.ea.2005}. However, possible role of magnetic nanoparticles in molecular processes of DNA degradation has not been discussed yet.

Exposure of humans to airborne nanosized particles has increased dramatically over the last decades due to anthropogenic sources. Rapidly developing nanotechnology provides yet another source of potential contamination. Nanoparticles can penetrate across epithelial and endothelial cells into the blood and lymph circulation to reach bone marrow, lymph nodes, spleen, and heart. Entry to the central nervous system and ganglia via translocation along nerve cells is also possible \cite{Oberdorster.ea.2005}. Biological activity of nanoparticles stems from their specific surface properties and includes inflammatory, pro-, and antioxidant potential. Therefore, the uncontrolled presence of nanoparticles in the environment is now considered as a potential threat for human health within the frames of \emph{nanotoxicology} \cite{Hardman.2006}. 

Among others, magnetic nanoparticles possess very special properties because of their relatively large magnetic moment. In other words, such particles are small magnets that behave like a compass needle. On the one hand, magnetic nanoparticles can rotate in an external MF, thus exerting a pressure on biological tissues, and, on the other hand, they produce their own and relatively large MF, i.e. endogenous MF, which in turn may affect magnetosensitive biochemical reactions. 

Magnetic nanoparticles may appear in an organism differently: (i) they can penetrate through the organism's surface as natural pollutants, especially as particles of iron oxides that abound in nature ; (ii) artificial magnetic nanoparticles may be introduced into an organism with certain aims and penetrate through cell membranes \cite{Rodriguez.ea.2005}; (iii) magnetic nanoparticles can appear inside biological tissues in the course of natural process of biomineralization.

Artificial magnetic nanoparticles are used in biology and medicine for different purposes. It is magnetic hyperthermia of tumors \cite{Stauffer.ea.1984,Hergt.ea.1998,Matsuoka.ea.2004,%
Pierre.ea.2005}, magnetic separation of proteins and cell sorting \cite{Saiyed.ea.2003,Safarik.ea.2004}, magnetically targeted drug delivery \cite{Hafeli.2004}, delivery of plasmid DNA into the bacterial cells \cite{Chen.ea.2006}, contrast enhancement in medical magnetic resonance imaging, and etc \cite{Gupta.ea.2005}.

Below we discuss in what way magnetic nanoparticles might be involved into the DNA degradation process. Estimates will be made for the MFs produced by the particles and average increment of the free radical content due to the enhanced MFs.

\begin{figure}[t]
\[ \includegraphics[width=0.66\textwidth]{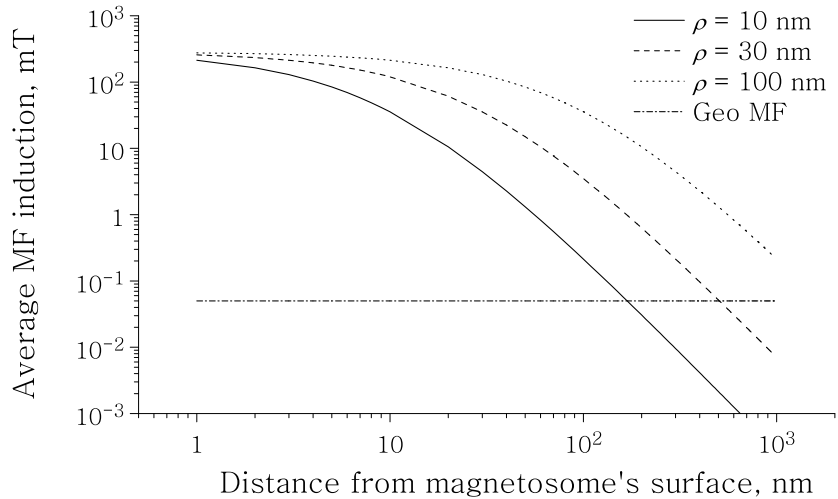} \]
\caption{\label{mfrad} {~~} {\protect \parbox[t]{100mm}{
Proper MF of idealized magnetosomes of different radii. Its value is far exceeds the level of the geomagnetic field.
} }}
\end{figure}

Magnetosomes have a magnetic moment and consequently they produce proper MF. This MF is not small, though it quickly drops with distance. An idealization is convenient, in which a magnetosome is a sphere of radius $\rho $ with a point magnetic moment $\boldsymbol \mu $ in its center. In the point ${\bf r} ={\bf n} r$, the magnetic moment generates MF $ {\bf H} = \left[ 3 {\bf n} (\boldsymbol \mu {\bf n}) - {\boldsymbol \mu} \right]/{r^3} $, where $\bf n$ is a unit vector in the direction of $\bf r$. MF vector $\bf H$ has direction and magnitude. The proper MF direction near the magnetosome changes quite considerably with its oscillations and may change to the opposite direction with the oscillation frequency. However, inasmuch as we consider MF effects on the rates of the radical-pair reactions, the MF direction is practically of little significance. The averaging of molecular orientations of a radical pair (RP) usually occurs far faster than the MF direction alters. Therefore, the rates of these reactions are sensitive only to the MF magnitude. Note that an anisotropic contribution to the rate of RP-reactions is theoretically considered in \cite{Ritz.ea.2004}, but there are still no reliable experimental verifications for this hypothesis. Then, we will study the averaged values of a constant and alternating components of the absolute MF value, or MF magnitude. 
The absolute MF value follows from the above expression for a dipole field:
\[ H(r,\theta , \varphi ) = \frac {\mu } {r^3} \sqrt{ 1+ 3\cos^2 \theta} ~, \]
where $\mu = | \boldsymbol \mu |$, and $r,\theta , \varphi $ are the radius vector and azimuth and polar angles of the spherical reference system, and the vector $\boldsymbol \mu$ is directed along the Z axis. 
It is easy to find the MF of a fixed dipole averaged over the volume between two imaginary spheres of radii $\rho $ and $R$:
\[ H_{\rm v} = \frac{c' 3\mu } {R^3-\rho ^3} \ln \frac R {\rho },
\]
where $c' = 1+ \ln (2+\sqrt 3)/(2\sqrt3) \approx 1.38$. From this we can readily find also the MF averaged over a sphere of radius $r$, if in the above formula for $H_{\rm v}$ we use the formal equalities $\rho = r$ and $R = r+{\rm d}r$, taking the limit as  ${\rm d}r \rightarrow 0$: $ \overline{H}(r) =   { \mu } c'/{r^3} $. Fig.~\ref{mfrad} demonstrates the dependence of the magnetosome's average MF $\overline{H}$, calculated by this formula, on the distance from the magnetosome surface. The fact by itself that the average proper MF of a magnetosome is many times greater than the geomagnetic field should result in some biological consequences.

The MFs produced by magnetosomes vary in a wide range, but mainly they fall within the interval 1--200~mT. Do such constant fields cause biological effects in general? Many studies, in which biological systems were exposed to MFs of these magnitudes, demonstrated various biological effects. In particular, such MFs can increase the level of the DNA damages \cite{Ho.ea.1992,Yokus.ea.2005,Saito.ea.2006}, deregulate the cell proliferative/apoptotic activity \cite{Fanelli.ea.1999,Robison.ea.2002,Ghibelli.ea.2006}, affect immunity \cite{Jandova.ea.2005} and gene expression \cite{Tokalov.ea.2004,Hirai.ea.2005}. 
Work \cite{Potenza.ea.2004} reports on mutagenicity of MFs, tumorigenicity of MFs was assumed in \cite{McCann.ea.2000,Thun-Battersby.ea.1999}, and MFs inhibitory activity with regard to apoptosis and tumor suppressor proteins was observed in \cite{Teodori.ea.2002,Teodori.ea.2005}. 

It is not infrequently argued that these effects might explain the correlation between childhood leukaemia incidence and the enhanced background extremely-low-frequency magnetic fields at places of residence \cite{Santini.ea.2005,Juutilainen.ea.2006}.

A number of works report no effect of static MFs on organisms. It is difficult to reconcile the above observations of different biological effects from static MFs with known reviews which claim absolute safety of medical magnetic resonance protocols \cite{Villa.ea.1991,Schenck.2005}. The magnetic resonance studies may really be safe due to their relatively short magnetic exposures. However, this does not rule out the potential health hazard associated with chronic MF exposures and  cumulative effects at the level of DNA mutations. Data are often contradictory: a part of studies report clear mutagenic, co-mutagenic, or toxic effects from tesla-range static MFs \cite{McCann.ea.1998,Ikehata.ea.1999,%
Takashima.ea.2004,Miyakoshi.2006}, while others do not \cite{Schreiber.ea.2001,Zhang.ea.2006}. 

At least in part, this discrepancy may originate from backward influence of the genome on the magnetic effects: different strains of the same animals displayed significantly distinct sensitivity to co-carcinogenic or tumor-promoting effects of MF exposure \cite{Fedrowitz.ea.2006} and different strains of the same bacteria showed clearly distinct frequency spectra of a magnetic effect \cite{Alipov.ea.1996}. Genes implicated in genetic instability syndromes were relevant in modulating the response of cells to an ac MF \cite{Mangiacasale.ea.2001}. To add, the available epidemiological data are not sufficient to draw any conclusions about potential health effects of static MF exposure \cite{Feychting.2005}. In this regard, a care should be taken: a negative result says either no effect takes place at all or the effect is less than the resolution of the experimental technique, \emph{a priori} half by half. Usually, accuracy of measurements in biology is less than a few percent even for large statistical samples due to normally great inherent variability of biological objects. But a few percent effects are extremely important when considered with regard to human populations.

Unlike constant fields, variable MFs induce electric currents in tissues, which may exceed natural biological currents, depending on the size and geometry of a biological system. It occurs when the frequency-amplitude product of an MF is greater than 10~Hz$\times $mT in order of value. For example, observation of a biological effect from 50~Hz MFs greater than 0.1~mT could not be followed by a statement that MFs of that magnitude entail the similar effects, because the actual cause might be in the electric fields. 

Particularly heavy for interpretation are experiments, in which a biological system is exposed to an intermittent or pulsed MF. Within the short time intervals, when the MF quickly changes its magnitude, great electric current pulses appear in biological tissues, which undoubtedly causes significant electrophoretic-like biological effects. Here, the MF is not a prime cause of the effect, but just a method for inducing a great electric current in tissue, like in transcranial magnetic stimulation.

For these reasons, all over the present article, we focus on the experiments, in which biological effects were produced by the constant MFs and extremely-low-frequency MFs lower than about 0.1~mT only. The following Table~\ref{tee02} summarizes the results of some recent experiments, sorted by the value of MFs used.

\begin{table}
\small
\caption{\label{tee02} Effects of static MFs on biological processes}\vspace{1mm}
\begin{tabular}{p{49mm}|p{20mm}|p{10mm}|p{9mm}|p{61mm}}
\hline
Object & B  & Mode$^{*}$ & Time & Effect$^{**}$ /Ref \\
       \hline\hline

Cell-free myosin phosphorylation & 0.1--180 $\mu $T & W & 60--90m & $140\pm 10$, \%  \newline\cite{Markov.2004} \\
                                 & 0--55 mT & M & & \\ \hline

DNA damage  & 8 $\mu $T, 50 Hz & W & 11--32d & +, $p<0.001$ \newline\cite{Svedenstal.ea.1999} \\ \hline

Single and double-strand DNA breaks  & 0.01 mT, 60 Hz & W & 1--2d & ~ \newline\cite{Lai.ea.2004} \\ \hline

Expression of heat shock genes in human leukaemia cells   & 0.01--0.14 mT, 50 Hz& W & 30m & sign. changes \newline\cite{Tokalov.ea.2004} \\ \hline

Human T lymphocytes adherence & 0.1--10 mT &  &  & ~ \newline\cite{Jandova.ea.2005} \\ \hline

DNA repair rate  & 0.15 mT, 60 Hz& W & 4--24h & +, $p<0.001$ \newline\cite{Robison.ea.2002} \\ \hline

Blood velocity in capillaries in mice & 0.3--10 mT & M & 10m & $42$\%, $p<0.05$  \newline\cite{Xu.ea.2001} \\ \hline

Developmental abnormalities in \emph{Drosophila} larvae & 0.5--9 mT & W & 30m & times \newline\cite{Ho.ea.1992} \\ \hline

Induced apoptosis in human leukocyte cells U937, CEM & 0.6--6 mT & M & 3h & $-64 \pm 6$, \%  \newline\cite{Fanelli.ea.1999} \\ \hline

Mechanosensitive ion channels in \emph{E. coli} & 1.35 mT& W & 1m & 0--2100, \% \newline \cite{Dobson.ea.2002}  \\ \hline

Stress-induced apoptosis in human glioblastoma cells & 6 mT & M & 18--24h & $-80\pm 15$, \%  \newline\cite{Teodori.ea.2002} \\ \hline

Cell morphology, apoptosis, gene products, 
etc., in different cells & 6 mT & & 24--48h & sign. changes  \newline\cite{Dini.ea.2005} \\ \hline

Shape and other morphology of human glioblastoma cells & 8--300 mT & M & 3h & $140\pm 50$, \%  \newline\cite{Teodori.ea.2005} \\ \hline

Brain malonedialdehyde, 
 NO synthetase activity in rats & 40 mT &  & 30m/d 7d & +, $p<0.05$  \newline\cite{Xu.ea.2006} \\ \hline

Norepinephrine content in gastrocnemius muscle in rats & 128 mT & M & 1h/d 5d & 25\%, $p<0.05$  \newline\cite{Abdelmelek.ea.2006} \\ \hline

Insulin concentration, blood, biochemistry in pregnant rats & 128 mT & M & 13d & $56\pm 16$, \%  \newline\cite{Chater.ea.2006} \\ \hline

Increase in plasma NO metabolites in 
rats & 180 mT & M & 6w & +, $p<0.01$  \newline\cite{Okano.ea.2006} \\ \hline

Plasmid stability in \emph{E. coli} & 250 mT & M & 2h &  30\% \newline\cite{Potenza.ea.2004} \\ \hline
Size of neurosecretory neurons in pupae of mealworm \emph{Tenebrio molitor} & 320 mT & M & 8d & $27\pm 6$, \%  \newline\cite{Peric.ea.2006} \\ \hline
Teratogenic effects on developing fetuses in mice & 400 mT  & M & 1h & 480--870, \%  \newline\cite{Saito.ea.2006} \\ \hline
\end{tabular}
$^{*}$ {\small M --- magnet, W --- coils or wires} \\ 
{$^{**}$ {\small Max. value, approximately}} 
\end{table}

\begin{figure}[t]
\[  \includegraphics[width=0.7\textwidth]%
{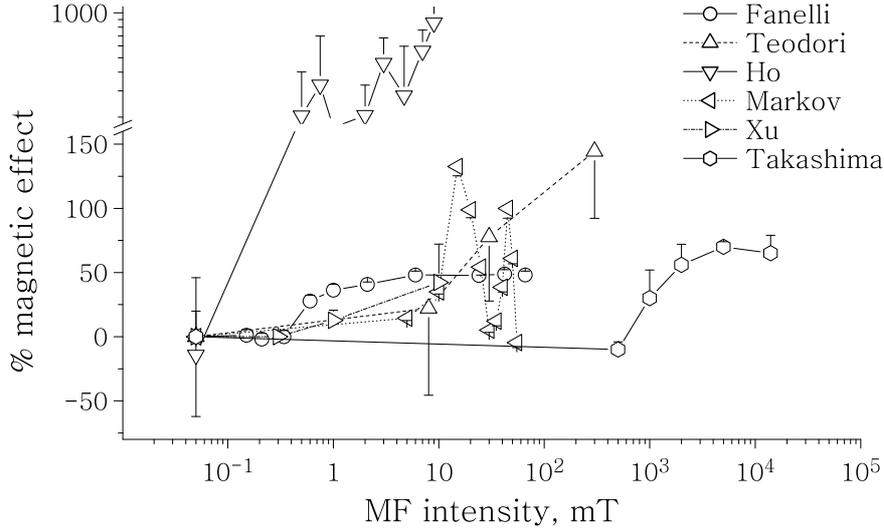} \]
\caption{\label{deg-SMF} {~~} {\protect \parbox[t]{100mm}{
Pooled data on the relative biological effect from exposure to static magnetic field, adapted from \protect\cite{Fanelli.ea.1999,Teodori.ea.2005,Ho.ea.1992,%
Markov.2004,Xu.ea.2001,Takashima.ea.2004}.
} }}
\end{figure}

As is seen, there are different effects of mT-range MFs on biological systems. Reviews \cite{Miyakoshi.2005,Dini.ea.2005} discuss a variety of such effects with more details and in a wider range of MF intensities. Just a few experiments investigated the field dependence of magnetic effects. These experiments can provide information on the physical nature of the MF targets.  Fig.~\ref{deg-SMF} shows pooled MF dependences of the relative magnetic effect in different biological systems, in which the experimental data were normalized in a similar way. As a general motif of Fig.~\ref{deg-SMF}, one could assume that the relative biological effect is approximately proportional to the MF magnitude, reaching about 20--40\% in 100~mT MFs.

On the whole, the MF dependences of this type resemble those for the rate of magnetosensitive chemical reactions involving spin-correlated radical pairs. Radical pair mechanism would be a nice candidate to explain biological effects of weak geomagnetic-like MFs, unless some principal difficulties limited its sensitivity by maximum values of about 1\%/mT for the majority of radical pairs \cite{Salikhov.ea.1984,steiner89}. However, this limitation does not matter at all, regarding observation of biological effects from MFs significantly greater than the geomagnetic field $\sim 0.05$~mT. There are experimental indications that radical pairs take part in primary interactions of mT-range MFs with biological systems \cite{Rollwitz.ea.2004,Timmel.ea.2004,Liu.Theil.2005,%
Yokus.ea.2005}. Melatonin, known free radical scavenger, suppressed DNA damage induced by a static MF in \cite{Jajte.ea.2001} and by a relatively weak 60-Hz MF in \cite{Lai.ea.1997}. Other free radical scavengers blocked magnetic-field-induced DNA strand breaks in \cite{Lai.ea.2004}; authors proposed that magnetic fields initiate an iron-dependent free radical generation process in cells, which can lead to genotoxic changes. The association between mT-range MFs and excessive production of free radicals is a robust hypothesis, since today no other equally elaborated idea exists of how such static MFs could affect biological processes. High priority is assigned by WHO to studies of a co-mutagenic effect concerning the carcinogenic potential of static magnetic fields with regard to RP reactions \cite{WHO.2006}. Thus, several reasons exist and many researchers are apt to think that biological effects from mT-range MFs, which vary roughly as the MF value, with a saturation in higher fields, originate from magnetochemical reactions with radical pairs, most likely highly reactive oxygen species \cite{Simko.ea.2004,Simko.2004}.

One of the possible sources of a chronic MF exposure is the enhanced content of magnetic nanoparticles in human tissues. 
Magnetic nanoparticles generate mT-range MFs and affect production of free radicals. The question is, how much? Let an idealized magnetosensitive reaction be ${\rm AB} \rightleftharpoons \dot{\rm A}\dot{\rm B} \rightleftharpoons \dot{\rm A} + \dot{\rm B}$, where the intermediate $\dot{\rm A}\dot{\rm B}$ is a spin-correlated RP in a \qqe{cage} formed by the molecules of the surrounding viscous medium.

One of the results of magnetochemistry is that the relative rate of recombination $\dot{\rm A}\dot{\rm B} \rightarrow {\rm AB}$ may change between 1 and 1/4, in some units, depending on the state of spin evolutions in the RP. An MF may change the rate just within this limits. Usually, magnetic field dependences in magnetochemistry are complex dependences that may contain minima and maxima. However, as a whole, the magnetic effects grow with MF. As a very rough approximation, we assume that the rate of free radical formation $K$, which is bound to the recombination rate, depends on the value of magnetic field $H$ in the following way
\begin{equation}\label{97} K = {K_0} \left( 1+ 3 \frac {H} {H_0 +H} \right) ~, \end{equation}
where $K_0$ is a minimum rate at $H=0$ and $H_0 $ is a characteristic MF, which lies between 10~mT and 1~T for most RP reactions. The approximation is justified also due to a lot of different RP reactions in organisms, which are poorly identifiable in experiments on MF effects. We note, that the above formula is an optimistic one: usually, just 1--2\%-relative changes are observed even in strong MFs. However, looking at Fig.~\ref{deg-SMF}, one could assume that a few percent changes in free radical content should lead to about ten fold higher alterations in biological observables, probably due to an interplay between nonlinear space- and time-scaled biochemical reactions. 

\begin{figure}[t]
\[ \includegraphics[width=0.7\textwidth]{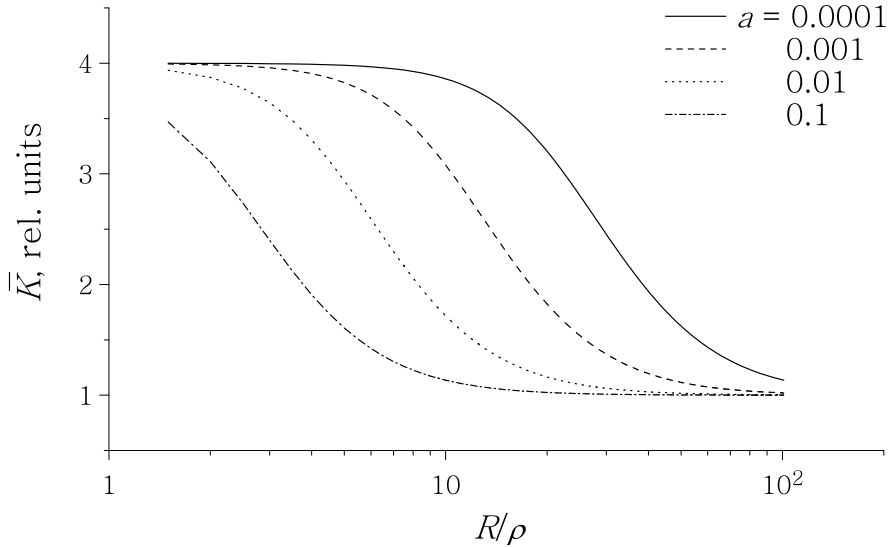} \]
\caption{\label{K-averaged} {~~} {\protect \parbox[t]{100mm}{
The average values of the rate of an idealized RP reaction in a volume of radius $R$ surrounding a magnetosome of radius $\rho $.
} }}
\end{figure}

Averaged value of $K$ over a volume of radius $R$ surrounding a magnetosome of radius $\rho $ is of interest. In calculations, the averaged MF of a magnetosome $\overline{H} \approx \mu \sqrt2 /r^3 $ is taken for $H$ and the magnetic moment $\mu = vJ$, where $v$ and $J$ are the magnetosome's volume and the saturation magnetization of magnetite Fe$_3$O$_4$. Averaged value of $K$, in $K_0$ units, is then as follows
\begin{equation}\label{deg100} \overline {K} \sim 1 + \frac 3 {a (R^3/\rho ^3 -1)} \ln \frac {1+a R^3 /\rho ^3} {1+a}  ~, ~~~ a \equiv 3H_0 /(4\pi \sqrt2 J) , \end{equation} This quantity is plotted on Fig.~\ref{K-averaged} as a function of $R/\rho$ at different values of $a$. As is seen, the average rate of the reaction markedly exceeds the level 1, which corresponds to the absence of the MF, only within the space limited by about ten-fold magnetosome radius.

It is known that for the majority of RP reactions, at the RP lifetime $\tau $ of the order of $10^{-9}$~s, the characteristic field $H_0$ is about 100~mT in the order of value. Correspondingly, the value of the parameter $a$ in (\ref{deg100}) equals about 0.1. For estimations, an inverse proportion between $\tau $ and $H_0$ may be assumed, so that relatively long lifetime $\tau \sim 1$~$\mu $s means that $a$ equals $10^{-4}$. However, so large lifetime could be an exceptional case: about 20~$\mu $s for the flavin--tryptophan \cite{Hore.ea.1987,Gindt.ea.1999} and 2--10~$\mu $s for flavin--tyrosyl \cite{Weber.ea.2002}. Therefore, it would be reasonable to assume $a \sim 0.1$, and thus the rate of most RP reactions might be shifted significantly only near magnetosomes, within the distance $R\sim \rho $ from magnetosomes' surface. At the same time, the presence of magnetic nanoparticles means a chronic magnetic exposure. In this case, as was shown above, even a few percent changes in the rate of RPs formation might be biologically significant due to accumulation of their contribution to DNA degradation. 5-\% changes in $\overline{K}$ take place over as large  volume as of the radius $R \sim 10$--$20$ $\rho $ for the straightforward case $a=0.1$ and $R \sim 100$--$200$ $\rho $ for $a=10^{-4}$. This unequivocally indicates that if a cell contains a magnetic particle, even as small as of a few nanometers, a significant part of the cell is covered by the MF of that particle, so that the rate of RP formation grows by about a few percent.

Probably, there is a hidden association between magnetic nanoparticles, which are composed of iron oxides mainly, and iron ions in cytoplasm that were shown to significantly increase the number of rat lymphocytes with DNA damage when administered simultaneously with the exposure to static MF of 7~mT \cite{Jajte.ea.2002}. The role of iron ions in free radical production is discussed in \cite{Lai.ea.2004}.

There are some proteins, like ferritin accumulating several thousands of iron atoms, which are necessary for biogenic magnetite production from ferrihydrite they contain \cite{Liu.ea.2005,Gossuin.ea.2005}. Iron compounds related to magnetite and ferritin were observed in a tumor tissue in \cite{Brem.ea.2005}. Nanoscopic magnetic crystals in ferritin cores may be a byproduct of biochemical processes utilizing Fe ions \cite{Quintana.ea.2004}. One might hypothesize that another yet unidentified proteins control and impede magnetite production, since its presence in extra amounts is harmful or toxic. Then, an equilibrium size of the magnetite byproducts should exist, since counteracting stimulus are to be proportionate to the surface of magnetite substance, which grows with size slower than mass. Probably, the size is very small and concentration of magnetic particles is extremely low under normal conditions, so that no biological consequences are possible. However, the dynamical equilibrium may be disturbed by the absence or inactivity of the specific proteins. As well, the equilibrium may vary in different organisms by genetic reasons, resulting in a chronically increased or decreased mean free radical concentration and thus in a higher or lower rate of genetic degradation and different effects on cancer probability. 

Time-course of the formation of extracellular biogenic magnetite was analyzed by quantum magnetic measurements in the culture of \emph{Geobacter metallireducens} \cite{Vali.ea.2004}. Authors of \cite{Scheffel.ea.2006}, using cryo-electron tomography that allows 3D reconstruction at rather high resolution, monitored the process of magnetosome formation in \emph{M. gryphiswaldense}. These studies have shown that transformation of ferrihydrite to magnetite is a complex many-stage process that can lead to formation of 5--10-nm crystallites within 30 min. Larger crystals grow during several weeks.  

Magnetic nanoparticles, depending on their size and substance, are in a multidomain, single-domain (SD), or superparamagnetic (SP) state. Biological formation of multidomain particles is seldom occurred. SD magnetite particles are of about 15--100~nm in size. They have received most attention in scientific literature as it is the most obvious and relatively easy to explain how they move mechanically, how external MFs act on their motion, and how this gives rise to activation of mechanoreceptors. The magnetic moment of an SD particle is rigidly bound to it, so that the MF interacting with the magnetic moment exerts a torque on the particle and mechanically rotates it.

Unlike SD particles, SP particles have their magnetic moments mostly unbound from their geometry. The magnetic moment of an SP particle can be switching between a few \qqe{easy} directions, which are determined by minima of the total magnetic energy. The latter includes the magnetic anisotropy energy of the bulk substance and demagnetizing energy that depends on the particle shape. Thermal perturbations make the particle to chaotically change its magnetic moment orientation, and no magnetic moment appears on average over time. An external MF brings the additional magnetic energy to the particle and makes it to preferably orient in only one of the easy directions. The time-averaged magnetic moment appears, however in weak MFs, it is very small as compared to the instant magnetic moment, and its magnetic energy is many orders of value lower than ${k_{\rm B}{\rm T}}  $. For this reason, no possible effects of SP particles have been studied thus far, except a few models where macroscopic conglomerates of such particles were considered from the viewpoint of their magnetic interaction with each other to explain magnetic navigation in some birds \cite{Davila.ea.2003}. 

In this article, an important role of single SP magnetic nanoparticles is emphasized. There are different ways in which they could magnetically influence biologically significant processes, in particular, the direct influence of the proper MF of magnetosomes on free radical formation, considered above. Another way is connected with the ability of SP nanoparticles to switch their magnetic moments time to time, being driven by random thermal forces. Frequency spectrum of this change-over depends mainly on the energy of the potential barriers, or maxima of the total magnetic energy. Since the switching is conjugated with a quick magnetic flux change, an electric eddy current, localized near the particle, is induced by this switching. On the one hand, this electric current should affect nearby chemical processes. On the other hand, this presents a dissipative factor for the switching process. Therefore, the magnetic switching, a non-inertial process in principle, is described by a dissipative dynamics with a nonlinear potential function. This is very similar to the stochastic dynamics of SD magnetosomes considered in \cite{Binhi.ea.2005b}. Different dynamic conditions are possible here, including stochastic resonance in external ac MFs and switching frequency control by static MFs.

The density of magnetosomes in the human brain is more than $5 \times 10^6$, and in meninges more than $10^8$ crystals per gram \cite{Kirschvink.ea.1992b}. In fact, about 90\% of the particles measured in this work were 10–-70nm in size, and 10\% were 90-–200 nm. Subsequent studies have shown that the concentration of magnetite/maghemite in human tissues varies from tens to hundreds ng/g \cite{Grassi.ea.1997} and equals about 50 ng/g on average in the human brain \cite{Schultheiss.ea.1999}. Magnetite levels may be even higher when considering superparamagnetic particles, diseased tissue and age \cite{Dobson.2002,Hautot.ea.2003}. No data are known about possible content of SP nanoparticles in human blood or marrow. Their possible contribution to magnetically measurable quantities could be masked by relatively large amounts of ferritin \cite{Brem.ea.2006} and diamagnetic and paramagnetic fraction of blood, which requires development of special methods for single SP nanoparticles detection.

The SP particles present a new and highly interesting object from the viewpoint both of quantum physics and of biology, particularly biology of cancer. Specific molecular mechanisms based on SP nanoparticle dynamics will be studied elsewhere.

\section{Discussion}

Enhanced cancer probability in the early period of life results from rapid growth in the number of cells in a population while the repair and immune systems are not yet ready to repair DNA damage effectively and remove the cells with mutant DNA molecules.

To be protected from early cancer, it is very important for an organism to form the DNA repair system and immune system as soon as possible, since even a small delay may result in a significant growth of risk. Therefore, exogenous factors may be dangerous even when they, per se, do not cause DNA lesions or mutations. If exogenous factors retard the formation of protective systems, they promote enhanced cancer risk. 

Formation of the immune system may go faster under training, i.e., during some slight viral diseases or other infections. Paradoxically, developing organisms need to be sick from time to time in order to be healthy on the whole. There is recent epidemiological evidence that any activity outside the home during the first year of life, especially the first three months, is associated with a decreased incidence of acute lymphoblastic leukaemia, by 40--50\% \cite{Gilham.ea.2005}; the most likely explanation being a protective effect from exposure to common infections. It is interesting that the mathematical model demonstrates this effect.

The other non-evident conclusion is that the risk of early leukaemias might be reduced by slowing down the division of cells of a growing organism in the first year(s) after impregnation in order to give more time to the immune system to develop. It is probable that a rough life for a woman during pregnancy and for her child after birth could promote an advanced development of immunity and decrease the probability of cancer in the first years. Children conceived in lean years may show a reduced cancer incidence rate in early age.

The DNA degradation model described here is a probabilistic model. Though valid with respect to cancer risk assessment for individuals, it could be verified only by epidemiological studies. However, the situation may change in the future. Assuming that the probability of cancer $P$ may be determined not only for a human population, as $P(t)$ in Fig.~\ref{P-deflection}, but for a separate human organism as well, one might speculate that there is an oncological age. Formally, it is the age $\tilde{t}$ where $P(\tilde{t})$ equals to the probability of cancer for that human being. Oncological age, unlike biological age and, moreover, physical age, can be significantly affected by long-term detrimental environmental conditions and short-term risk factors. Oncological age, very similar to Balzac's magic piece of shagreen, shows the probability of cancer for this human in terms of age. Most likely, the future success of nanotechnology will make it possible to straighten DNA molecules and read out genetic information quickly by physics methods. Then, oncological age will become a truly measurable quantity.

The model might be further corrected in many aspects. An obvious one is that the model has no specificity with regard to different genes. One more aspect is that there are different classes of DNA damage/mutations and they have different rates. Another aspect is that the model studies a population of idealized identical cells, while cells with parameters, distributed over some ranges, would be more realistic. For example, it would be right to assume that only some of the cells contain magnetic nanoparticles and then have a significantly increased rate of DNA degradation, rather than that all cells have a moderately increased rate. Yet another aspect is that there are \qqe{hot and cold spots} for spontaneous mutations, i.e. DNA sites that have a higher or lower rate of mutations than expected from a normal distribution. Apparently, the list might be continued. However, even in its current simple form, the model is effective. Basic clinical observations, like the exponentially growing probability of cancers in middle age and an enhancement of the probability for some cancers in early childhood, have a clear explanation within a modeling context.

Some of the results of the model are probably new and unobvious as mathematical corollaries: there is a synergistic effect of harmful factors, like free radicals, acting through the different biological/biochemical pathways or channels of DNA degradation, and there are unequal long-term consequences from the same short-term detrimental exposure applied in different phases of an organism's development.

A possible source of an increased level of free radicals is exposure to relatively strong MFs from magnetic nanoparticles in human tissues. If a cell contains a magnetic particle, even as small as a few nanometers across, a significant part of the cell is covered by the MF of that particle, so that the rate of free radical formation may grow by a few percent. Magnetic nanoparticles, chronically increasing local concentration of free radicals around themselves in cells, may affect cancer related processes through all the channels: increasing rate of lesions to DNA molecules, decreasing rate of DNA repair, impeding immune responses and apoptosis, causing acceleration of an organism's development and shortening cell cycles. Since, in general, biological effects of MFs are known not to be strong, it is reasonable to assume that a most sensitive system is a more probable target for MFs in living matter. Within the model context, it is the immune system. 

External ac MFs may change the dynamics of magnetic nanoparticles in different ways. By facilitating the switching of the particle magnetic moment and so the rate of electric pulses induced, ac MFs can increase the rate of free radical formation. By shaking magnetic moments, ac MFs can impede adaptation of cell structures to their presence. It can also shift the spectrum of the endogenous ac MFs, and unbalance time-ordered cycles of RP biochemical reactions. These specific mechanisms are hypothetical and require further estimates.

How to test whether SP nanoparticles contribute to genome degradation? The most direct way would be to observe a difference in the amount of the particles in tissues of cancer patients and healthy people. Another approach to identifying this specific risk factor is to search for an association between the amount of the particles and the age at diagnosis of a leukaemia. Some indirect indications may come from epidemiological studies. On the one hand, stochastic resonance of magnetosomes in ac MFs \cite{Binhi.ea.2005b} shows non-trivial static MF dependence of the induced biological effects. On the other hand, an increasing trend of childhood leukaemia were associated with certain combinations of household static and power-frequency MFs \cite{Bowman.ea.1995}. Laboratory studies could also be useful: SP nanoparticles may be artificially introduced into cell cultures or whole organisms, which could result in cancer development in chronic experiments.

\section{Conclusion}

In summary, a mathematical model has been developed that explains basic oncological regularities and offers unobvious strategies for minimizing risks of some cancers. It is proposed and substantiated that an enhanced level of acute lymphoblastic leukaemia in early childhood may originate from magnetic nanoparticles located in hematopoietic stem cells. The magnetic nanoparticles may have a natural biogenic origin from ferritin or they may appear in the cells due to exogenous contamination by widely spread iron oxide nanoparticles. Alternating magnetic fields interacting with these particles change their contribution to free radical biochemical processes, responsible for the development of the immune system. Excessive free radicals impede development of this system and delay its maturation, thus resulting in an enhanced leukaemia incidence.

\section*{Acknowledgments}

This work was funded by the UK charity, CHILDREN with LEUKAEMIA.

\section*{Addendum. Numerical values of the model parameters}

Human genes vary enormously in size and exon content. Exons are regions of genes that code for main portion of proteins. There is an inverse relationship between gene length and percentage of exon content, so that a great variety of different genes have exon lengths between 0.2 and 20~kb \cite{Strachan.ea.1999}. Apparently, only mutations to exons have detectable phenotypic effects, while others do not alter protein functions. Due to these facts, the parameter $k$, the number of bits in a gene is taken to be $k=2\times 10^3$, i.e. in the middle of the range 0.2--20~kb.

Mutations are induced in DNA by exposure to a variety of mutagenic factors in both intracellular and external environment. Spontaneous errors in DNA replication and repair are the main source of mutations. For higher eukaryotes, the rate of spontaneous mutation per generation is about one mutation per gamete \cite{Strachan.ea.1999}. The size of the human DNA sequence is near $3\times 10^9$~bp. So, the rate of mutations per bp, or bit, is about $p \sim 3\times 10^{-10}$. Since the coding DNA of an average human gene is 0.2--20~kb, mutations occur spontaneously with an average rate from about $6\times 10^{-8}$ to $6\times 10^{-6}$ per gene per cell division, which is in accordance with known estimates of the rate of spontaneous mutation \cite{Drake.ea.1998,Jackson.ea.1998,Drake.2006}. In the present model, the probability of mutation per bit per division $p\sim \alpha \nu $ that follows from (\ref{degrad02}). Since $\nu $ is a fitting parameter, found to be about $10^{-3}$~d to fit data on Fig.~\ref{P-largetimes}, parameter $\alpha $, the rate of lesions per bit per cell division, has been taken $10^{-6}$. 

HSCs enter the cell cycle with different rates, since \emph{in vivo} they are usually in different cell cycle phases. In mice, approximately 8\% of HSCs asynchronously entered the cell cycle per day and about 50\% in 6 days \cite{Cheshier.ea.1999}. Cell cycle parameters of murine bone marrow cells were estimated in \cite{Bernard.ea.2003} based on the comparison between a cell proliferation model \cite{Mackey.2001} and experimental data from cell tracking experiments. Cell cycle duration was found to be about 0.3~d. In our work, the cycle duration of HSCs is a function of time within the limits 0.5 to 10~d, obtained by a conversion from the tabulated time dependence of the mass of the human organism.

Reparation time $\tau $ describes the efficiency of the DNA repair system. The repair efficiency is assumed to depend on accumulated mutations. At the initial relatively long period of a slow growth of $P$, time $\tau $ is practically equal to its initial value $\nu $. Apparently, there are a wide spectrum of times needed to repair DNA damages. Normal cells react to an unrepaired DNA damage by retarding the cell cycle at a checkpoint until the damage is repaired, and triggering apoptosis when the damage is lethal. Retardation may continue from days to years. Proliferation kinetics of HSCs isolated from mice has been studied in \cite{Cheshier.ea.1999}: about 8\% of cells asynchronously entered the cell cycle per day, 75\% of cells were quiescent at any one time, and all cells were recruited into cycle regularly such that 99\% of them divided on average every 57 days. The values between 20 and 50 days were used in \cite{Mackey.2001} in modelling HSC kinetics.

It is unclear to which extent one could identify retardation time and repair time. Complex \qqe{digital} interaction between repair process, cell cycle arrest and apoptosis has been described in \cite{Ma.ea.2005}. Two character times, fast and slow, of the double-strand break repair process were suggested in \cite{Stewart.2001}; most of the breaks are rejoined quickly for 15 min and the rest are rejoined very slowly, for 10 to 15 h repair half-time. In the present model, cells enter the next cycle without delay, and repair time is a separate intracellular process. The value of repair time $\nu $ was taken on the order of 0.001~d, i.e. about a minute, to get probability curves fitted well to epidemiological data, Fig.~\ref{P-largetimes}. This means, in this model, parameters $\nu $ and $\tau $ should not be deemed precisely as repair time, it is rather parameters describing overall efficiency of the repair system.


\end{sloppypar}
\end{document}